\documentclass[aps,prl,showpacs,twocolumn]{revtex4-1}
\usepackage[utf8x]{inputenc}
\usepackage{amssymb}
\usepackage{graphicx}
\usepackage{color}
\begin{document}
\title{Fermi liquid theory and ferromagnetism} 
\author{V. P. Mineev}
\affiliation{Commissariat \`a l'Energie Atomique,
INAC/SPSMS, 38054 Grenoble, France}
\date{\today}

\begin{abstract}
There is demonstrated that 
an isotropic ferromagnetic Fermi liquid reveals instability of the ferromagnetic state in respect to the transversal inhomogeneous deviations of magnetization from equilibrium.  The result was obtained by derivation of the  spin waves spectrum  by means of kinetic equation.
\end{abstract}


\maketitle

{\it Introduction.}

Ferromagnetism in a system of itinerant fermions is traditionally explained  in terms of physical representations first introduced by Stoner
\cite{Stoner}. It arises  due to  the short range repulsion between the particles with the opposite spins 
favoring the finite magnetic polarization hindered by the kinetic energy grown  caused by the Pauli exclusion principle. This physical idea has found  more exact  formulation  in frame of the Fermi gas model \cite{Huang}. After appearance of the  Landau Fermi liquid theory \cite{Landau} there was developed the corresponding description of ferromagnetic Fermi liquid \cite{AbrDz,DzyalKond1976}. 
  The fast development of cold gases physics has animated
the interest to the problem of itinerant ferromagnetism \cite{Jo2009}. The Stoner model has started to be the subject of more careful investigations.

Still, theoretically there is the question:
can a gas of spin-up and spin-down fermions  with pure repulsive short range interaction be ferromagnetic?  
On the one hand there is rigorous statement  \cite{Barth2011} based on the Tan relations that in both one and three dimensions, a Stoner instability to a saturated ferromagnet for repulsive fermions with zero range interactions is ruled out at any finite coupling. However, it rules out only {\it saturated} ferromagnetism for the {\it homogeneous} gas but not partially polarized states. The condition of the zero range interaction seems  to be also quite restrictive. 

On the other hand the problem of transition to the ferromagnetic state in a two-component repulsive Fermi gas has been addressed in many studies (see \cite{Huang, Kanno,Belitz,Heiselberg2001,Duine2005, Conduit,Pilati2010,Chang,Heiselberg2011,He2011}.
 In particular, there were established that in the first order perturbation theory \cite{Huang} the ferromagnetic phase transition is of the second order and occurs at $k_Fa=\pi/2$ whereas the second order perturbation theory \cite{Duine2005} predicts the first order phase transition at $k_Fa=1.054$. Finally nonperturbative study \cite{He2011} returns us back to the phase transition of the second order at $k_Fa=0.858$.
These treatments  were also performed  under the assumption that only phases with 
{\it homogeneous} magnetization can be produced. So, the properties of the Stoner-Hubbard model still are under active theoretical investigation.

Leaving apart the problem of a Fermi liquid phase transition to the ferromagnetic state here we discuss the stability of  this state in respect to the transversal inhomogenious magnetization perturbations.
The transverse spin wave dispersion in a ferromagnetic Fermi liquid  at $T=0$  in the assumption of absence of the quasiparticle excitations  attenuation has been obtained by Abrikosov and Dzyaloshinskii \cite{AbrDz}. 
There was calculated 
the spin waves spectrum $\omega=D^{\prime\prime}k^2$ and  the reactive part of diffusion constant $D^{\prime\prime}$ was found proportional to the spontaneous magnetization. The theory has been criticized by C.Herring \cite{Herring} who has remarked, that in a ferromagnetic Fermi liquid the quasiparticle states with, say, spin up will no longer be closed
to the Fermi surface of quasiparticles with the opposite spins. So, they "will have a finite, rather than an infinitesimal, decay rate".  In its turn, the spin wave  dispersion  
has to  require an imaginary part: $D^{\prime\prime}\to-i(D^{\prime}+iD^{\prime\prime})$.
Later the reactive part of spin wave spectrum  has been calculated by Moriya \cite{Moriya} in frame of RPA approach applied to the ferromagnetic Fermi gas with  Hubbard type interaction.  Again the spin wave attenuation was not found.
The problem was addressed
in the paper by the author\cite{Mineev2005}. 

\bigskip

{\it Kinetic equation approach.}
The dispersion law 
\begin{equation}
\omega=\omega_{L}+ (D^{\prime\prime}-iD^{\prime})k^{2}
\end{equation}
for transversal spin waves was found for  Fermi liquid
both in paramagnetic and ferromagnetic states.
Here $\omega_L=\gamma H_0$ is the Larmor frequency, $H_0$ is the external field, $\gamma$ is the gyromagnetic ratio. To escape to look after $\gamma$ and $|\gamma|$ we treat $\gamma$ as a positive constant.
The dispersion law is derived making use the Silin kinetic equation \cite{Silin} under conditions that the dispersive part is  much smaller than the scattering rate, the latter in its turn is much smaller than the liquid polarization $\gamma H$ assumed small in comparison with the Fermi energy \cite{Mineev2005} 
\begin{equation}
|D|k^2<<\frac{1}{\tau}<<\gamma H.
\label{cond}
\end{equation}
The polarization  in the paramagnetic state can be created either by the external field or by the pumping.
In the former case it is determined by
the external field and the Landau molecular field parameter $F_0^a>-1$
\begin{equation}
\gamma H=\frac{\gamma H_0}{1+F^a_0},
\end{equation}
whereas in the ferromagnetic state, when $F_0^a<-1$, it has the finite value 
\begin{equation}
\gamma H=4\sqrt{6}~\varepsilon_F\sqrt{\frac{1+F_{0}^{a}}
{F_{0}^{a}}}
\end{equation}
even in the  external field  or the pumping absence. The corresponding magnetic moment density is given by
\begin{equation}
{\bf M}=
\frac{\gamma^2 N_{0}}{4}{\bf H}.
\end{equation}
Here $N_0$ is the quasiparticles density of states at the Fermi surface. We put $\hbar=1$ throughout the paper. 

The quasiparticles scattering rate in the polarized Fermi liquid is \cite{Min}
\begin{equation}
\frac{1}{\tau}=A((2\pi  T)^2+(\gamma H)^2),    
\end{equation}
Here, $A =const\propto m^{*3}W  $, $m^{* }$ is the effective mass and $W$ is square modulus of the matrix element of the opposite spin quasiparticles short range potential of interaction.  
The scattering rate has a finite value $\frac{1}{\tau}\propto(\gamma H)^2$ even at $T=0$. This statement
for spin polarized paramagnetic Fermi liquid has been checked and confirmed experimentally \cite{Perisanu}.

Under fulfillment conditions (\ref{cond}) the reactive $D^{\prime\prime}$ and the dissipative $D^{\prime}$
parts of the diffusion constant are given by the following expressions \cite{Mineev2005}
obtained taking into account  the amplitude  $F_1^a$ of the first angular harmonic of the quasiparticles exchange interaction.

\bigskip

%\begin{widetext}
$~~~~~~~~~~~~~~~~~~~~~~~~~~~~~~~~$\tablename  
\bigskip
\\
\begin{tabular}{|c|c|}
\hline
Polarized paramagnetic&Ferromagnetic at $H_0=0$\\
\hline
$D^{\prime\prime}=\frac{v_F^2(1+F_0^a)\left(1+\frac{F_1^a}{3}\right)}{3\left(F_0^a-\frac{F_1^a}{3}\right)\gamma H}$&
$
D^{\prime\prime}=\frac{v_F^2F_0^a\left(1+\frac{F_1^a}{3}\right)\gamma H}{3(4\varepsilon_F)^2\left(F_0^a-\frac{F_1^a}{3}\right)}>0$\\
$
|D^{\prime\prime}|\propto\frac{1}{H}
$&$
D^{\prime\prime}\propto H
$\\
\hline
$
D^{\prime}=\frac{v_F^2(1+F_o^a)\left(1+\frac{F_1^a}{3}\right)}{3\left(F_0^a-\frac{F_1^a}{3}\right)^2(\gamma H)^2\tau}>0
$&
$
D^{\prime}=-\frac{v_F^2(F_0^a)^2\left(1+\frac{F_1^a}{3}\right)}{3(4\varepsilon_F)^2\left(F_0^a-\frac{F_1^a}{3}\right)^2\tau}<0$\\
$
D^{\prime}\propto const
$&$
|D^{\prime}|\propto H^2
$\\
\hline
\end{tabular}
%\end{widetext}

\bigskip

The reactive part of diffusion coefficient $D^{\prime\prime}$ in ferromagnetic Fermi liquid derived from kinetic equation 
 \cite{Mineev2005} literally coincides  with result of Moriya \cite{Moriya} obtained in frame of RPA  at $F_1^a=0$ applied to the Hubbard model. One can prove also the exact correspondence of these two approaches for the $D^{\prime\prime}$ in polarized paramagnetic Fermi liquid.
  
 Sign of coefficient $D^{\prime\prime}$ in polarized paramagnetic state can be positive or negative depending on sign of the difference $F_0^a-\frac{F_1^a}{3}$.  In ferromagnetic state where $F_0^a<-1$, and $F_0^a-\frac{F_1^a}{3}<0$ the sign of coefficient $D^{\prime\prime}$ 
 is always positive.
 
The dissipative parts $D^{\prime}$ of the diffusion coefficients in the polarized paramagnetic Fermi liquid and in the ferromagnetic Fermi liquid have the opposite signs. As result the amplitude of transversal inhomogeneous deviations of magnetization
\begin{equation}
\delta M\propto \exp(ikx-i\omega t)=\exp(ikx-iD^{\prime\prime}k^2t-D^{\prime}k^{2}t)
\end{equation}
attenuates in a polarized paramagnetic Fermi liquid   but it grows up in a ferromagnetic Fermi liquid
demonstrating instability of the Stoner state in respect of such deviations.

\bigskip

{\it Field theory approach.}
 The calculation of reactive part in ferromagnetic Fermi liquid was performed by Dzyaloshinskii and Kondratenko \cite{DzyalKond1976}.  They have note that  the amplitude of forward scattering \cite{Landau1958} 
for the quasiparticles with the opposite spins  is proportional to the static transverse susceptibility 
which is divergent
in a isotropic ferromagnet. 
 In application to the
 spin polarized paramagnetic Fermi liquid one must use the divergence of transverse susceptibility at 
 frequency equal to the Larmor frequency.
 Following this schema and  taking into account the imaginary self-energy of quasiparticles \cite{Mineev2005} one can obtain  the dispersion laws for ferromagnetic Fermi liquid
 \begin{equation}  
\omega\approx\left[\gamma H-ic(\gamma H)^2\right]\frac{k^{2}}{p_F^2},
\end{equation}
and for paramagnetic Fermi liquid
\begin{equation}  
\omega-\omega_L\approx\left[-(\gamma H)^{-1}+ic\right]{v_F^2}k^2.
\end{equation}
Here,  $c$ is positive constant. For dilute Fermi systems 
$
c\propto m^*\sigma
$ (in dimensional units $c\propto m^*\sigma/\hbar^2$ ), where $\sigma$ is the cross-section for the s-wave scattering of particles with opposite spins. 

 We see: both in paramagnetic and ferromagnetic Fermi liquid cases the polarization dependence and the sign of the reactive part of spin diffusion coefficient $D^{\prime\prime}$ derived by the field theoretical method 
are in correspondence with kinetic equation results. As for the dissipative part $D^{\prime}$ a correspondence is absent. In contrast to the kinetic equation approach here we have : 
$D^{\prime}>0$ for ferromagnetic Fermi liquid  and $D^{\prime}<0$ for the polarized paramagnetic one.  It is difficult to consider the latter result as satisfactory because
 phenomenological kinetic theory are confirmed by the microscopic derivation of kinetic equation for the  paramagnetic Fermi gas  with repulsion by means of the Keldysh technique \cite{Gol}.
In our opinion it means that the method developed in the paper \cite{DzyalKond1976} is appropriate for the calculation of reactive part of thansverse spin wave dispersion  but it is inapplicable to calculation of spin wave dissipation. \newpage

{\it Conclusion}

The problem of existence of itinerant ferromagnetism in system of Fermi particles with repulsive interaction was recently addressed in the experiments with ultracold two component Fermi gas
( $^6$Li atoms in the lowest two hyperfine states \cite{Jo2009}).
By the increasing of magnetic field toward the Feshbach resonance \cite{Chin}  one can reach the pairing instability  that is  formation of 
stable gaseous phase consisting of two opposite spin atom molecules.  However, the molecule production is slow, as it requires three-body process.
So, one can expect that   fast enough field magnification prepares  the system state characterized by  
 large ( comparable to inter-particle distance) opposite spin atoms scattering length $a>0$ formally corresponding to the strong
short range repulsion that
 leads to the formation of ferromagnetic state.
The latter is certainly metastable but one can hope that at fast nonadiabatic increase of scattering amplitude the ferromagnetic state will be formed faster than more stable gas of Li$_2$ molecules.
 
The observable experimental signatures of cloud of fermions in the harmonic optical  trap as a function of interaction 
has been calculated \cite{LeBlanc}. 
They are characterized by the maximum of cloud size and minimum of kinetic energy at phase transition to the ferromagnet state.
The corresponding nonmonotoneous dependences 
 has been measured \cite{Jo2009} but no magnetization grown has been resolved.
Later there was demonstrated, that the pairing instability  is always stronger than the ferromagnetic one \cite{Pekker2011}. 
Moreover, there was found theoretically  \cite{Zhang2011} and then confirmed experimentally \cite{Sanner2011} that the maximum atom loss at a magnetic field below Feshbach resonance is the result of processes of atomic dimers formation and relaxation. 
The authors of \cite{Sanner2011} conclude that unlike to the previous
 observations \cite{Jo2009}
"$\dots$ the sample remains in the paramagnetic phase for a wide range in the strength of interactions and wait time."

So, the cold atomic gases do not undergo a ferromagnetic phase transition. Being disappointed by their findings the authors formulate even more stronger statement: {\it "$\dots$ the Nature does not realize a strongly repulsive Fermi gas with short range interaction, and the widely used Stoner model is unphysical."} 

Indeed, one cannot point out a real ferromagnetic system  described by the Stoner model. As an  apparent exception could be the Quantum Hall Ferromagnetic state  formed according to terminology accepted in the experimental publications \cite{Piot,Zhuravlev} as result "a magnetic-field-induced Stoner transition". In fact the mechanism forming this state has nothing common with the short range Stoner-Hubbard 
 repulsion between the particles with antiparallel spin. On the contrary, it originates from the long range attractive exchange interaction between the electrons with parallel spins (for review see S.M.Girvin \cite{Girvin}) first introduced in application to the itinerant electron systems 
  by Felix Bloch \cite{FBloch}.

Here we have demonstrated the intrinsic instability in respect to the transversal inhomogeneous magnetization perturbation of an isotropic itinerant ferromagnetic state described in terms of the Fermi liquid theory
valid for the systems with short range interparticle repulsion.
The result  is obtained by means of derivation of dispersion of the transversal spin waves making use the Silin  kinetic equation. 
Due to the instability of an isotropic itinerant ferromagnetic state one can conclude even stronger than 
that was done in the paper \cite{Sanner2011}: {\it  if  the Nature does  realize a strongly repulsive Fermi gas with short range interaction it does not lead to the formation of ferromagnet state.}

As for ferromagnetic  metals there are a lot properties differing   real itinerant ferromagnets from the isotropic model describing Fermi gas with short range repulsion between the  particles. 
The simplest distinction is that in ferromagnetic metals the two-moment approximation does not work. The quasiparticle exchange interaction should be expanded not by the Legendre polynomials 
      but by the eigen functions of the irreducible representation of the point crystal symmetry group taking into account spin-orbital interaction.  The ferromagnetism in metals should be treated taking into account semi-itinerant nature of quasiparticles and the anisotropy of the quasiparticles interaction originating from the spin-orbital effects.

%      Example: crystal with cylindrical (hexagonal) symmetry:
%      $$
%f_{{\bf k}{\bf k'}}^{\sigma \sigma'}=
%f_{{\bf k}{\bf k'}}^{s}\hat I\hat I'+
%[f_{0}^{a}+f_{1}^{a}(\hat{\bf k}\hat{\bf k'})] \mbox{\boldmath$\hat \sigma
%$}\mbox{\boldmath$\hat \sigma' $}.
%$$
%$$
%f_{{\bf k}{\bf k'}}^{\sigma \sigma'}=f_\perp^a(\sigma_xk_x+\sigma_yk_y)(\sigma'_xk'_x+\sigma'_yk'_y)+f_\parallel^a(\sigma_zk_z)(\sigma'_zk'_z)+ \dots
%$$


\begin{thebibliography}{99}

\bibitem{Stoner} E. Stoner, Philos. Mag. {\bf 15}, 1018 (1933).

\bibitem{Huang} K. Huang, {\it Statistical mechanics}, Wiley, New York, 1987.



\bibitem{Landau} L.D.Landau, ZhETF {\bf 30}, 1058 (1956) [Soviet Phys. JETP {\bf 3}, 920 (1956)];
ZhETF {\bf 32}, 59 (1957) [Soviet Phys. JETP {\bf 5}, 101 (1957)]

\bibitem{AbrDz}A. A. Abrikosov and
I. E. Dzyaloshinskii, Zh.  Eksp.Teor.Fiz {\bf 35}, 771 (1958)
[Sov.Phys.JETP {\bf 8}, 535 (1958)]. 

\bibitem{DzyalKond1976}I. E. Dzyaloshinskii, P. S. Kondratenko, Zh.  Eksp.Teor.Fiz. 
{\bf 70}, 1987 (1976) [Sov.Phys.JETP {\bf 43}, 1036 (1976)].

\bibitem{Jo2009}G.-B. Jo, Y.-R. Lee, J.-H. Choi, C. A. Christensen, T. H. Kim, J. H. Thywissen, D. E. Pritchard, and W. Ketterle, Science {\bf 325}, 1521 (2009).


\bibitem{Barth2011}M. Barth, W. Zwerger, arXiv: 1101.5594.



\bibitem{Kanno} S. Kanno, Progr. Theor. Phys. {\bf 44}, 813 (1970).

\bibitem{Belitz} D. Belitz, T. R. Kirkpatrick,  T. Vojta, Phys. Rev. Lett. {\bf 82}, 4707 (1999).

\bibitem{Heiselberg2001} H. Heiselberg,Phys. Rev. A {\bf 63}, 043606 (2001).

\bibitem{Duine2005}R. A. Duine and A. H. MacDonald, Phys. Rev. Lett. {\bf 95}, 230403 (2005).

\bibitem{Conduit} G .J. Conduit,  B. D. Simons, Phys. Rev. A {\bf 79}, 053606 (2009);  G .J. Conduit et al, Phys. Rev. Lett.{\bf 103}, 207201 (2009).

\bibitem{Pilati2010}S. Pilati, G. Bertaina, S. Giorgini, and M. Troyer, Phys. Rev. Lett. {\bf 105}, 030405 (2010).

\bibitem{Chang} S.-Y. Chang et al, Proc. Nat. Acad. Sci. {\bf 108}, 51 (2011).

\bibitem{Heiselberg2011}H. Heiselberg,Phys. Rev. A {\bf 83}, 053635 (2011).

\bibitem{He2011} L. He and X.-G. Huang, arXiv:1106.1345


 

\bibitem{Herring}C. Herring "Exchange Interactions among Itinerant Electrons"
 Chapter XIV, pp.345-385, in "Magnetism" v.IV, edited by G. T. Rado and
 H. Suhl, Academic Press, NY and London, 1966.
 
\bibitem{Moriya} T. Moriya {\it Spin fluctuations in itinerant electron
magnetism} (Springer-Verlag, Berlin, 1985).

\bibitem{Mineev2005} V. P. Mineev, Phys. Rev. {\bf 72}, 144418 (2005).

\bibitem{Silin}V. P. Silin, Zh.  Eksp.Teor.Fiz.  {\bf 33}, 1227 (1957)
[Sov.  Phys.JETP {\bf 6}, 945 (1958)].

\bibitem{Min}V. P. Mineev, Phys.  Rev.  B {\bf 69}, 144429 (2004).

\bibitem{Perisanu}S. Perisanu and G. Vermeulen, Phys. Rev B {\bf 73} 214519 (2006).




\bibitem{Landau1958}L. D. Landau , Zh.  Eksp.Teor.Fiz.  {\bf 35}, 97 (1958)
[Sov. Phys. JETP {\bf 8}, 70 (1959)].

\bibitem{Gol}D. I. Golosov and A. E. Ruckenstein,
Phys.Rev.Lett.  {\bf 74}, 1613 (1995); Journ.  of Low Temp.Phys.  {\bf
112}, 265 (1998).


\bibitem{Chin} C. Chin, R. Grimm, P. Julien and E. Tiesinga, Rev. Mod. Phys. {\bf 82}, 1225 (2010).

\bibitem{LeBlanc} L. J. LeBlanc, J. H. Thywissen, A. A. Burkov, A. Paramekanti, Phys. Rev. A {\bf 80} 013607 (2009).


\bibitem{Pekker2011} D. Pekker, M. Babadi, R. Sensarma, N. Zinner, L. Pollet, M.W.Zwierlein, and E. Demler, Phys. Rev. Lett. {\bf 106}, 050402 (2011).

\bibitem{Zhang2011} S. Zhang and T.-L. Ho, New J. Phys. {bf 13}, 055003 (2011).

\bibitem{Sanner2011} C. Sanner, E. J. Su, W. Huang, A. Keshet,J. Gillen, and W. Ketterle, arXiv: 1108.2017.

\bibitem{Piot}B. A. Piot, D. K. Maude, M. Henini, Z. R. Wasilevski, K. J. Friedland, R. Hey, K. H. Ploog, A. I. Toropov, R. Airey and G. Hill, Phys. Rev. B {\bf 72}, 245325 (2005).

\bibitem{Zhuravlev} A. S. Zhuravlev, A. B. Van'kov, L. V. Kulik, I. V. Kukushkin, V. E. Kirpichev, S.H.Smet, K. v. Klitzing, V. Umansky, and W. Wegsheider, Phys. Rev. B {\bf 77}, 155404 (2008).



\bibitem{Girvin}S. M. Girvin, ArXiv: cond-mat / 9907002.

\bibitem{FBloch}F. Bloch, Z. Phys. {\bf 57}, 545 (1929).






\end{thebibliography}
\end{document}